\documentclass[aps,prl,twocolumn,superscriptaddress,showpacs]{revtex4}
\usepackage{graphicx}
\begin{document}
\def\T{\Theta}
\def\D{\Delta}
\def\d{\delta}
\def\r{\rho}
\def\p{\pi}
\def\a{\alpha}
\def\g{\gamma}
\def\ra{\rightarrow}
\def\s{\sigma}
\def\b{\beta}
\def\e{\epsilon}
\def\G{\Gamma}
\def\om{\omega}
\def\pe{$1/r^\a$ }
\def\l{\lambda}
\def\f{\phi}
\def\w{\psi}
\def\m{\mu}
\def\t{\tau}
\def\dg{d^3{\bf r}\,d^3{\bf p}}
\def\df{f({\bf r, p})}
\def\dn{n({\bf r, p})}

\title{Collapses and explosions in self-gravitating systems}
\author{I.~Ispolatov}
\affiliation{Departamento de Fisica, Universidad de Santiago de Chile,
Casilla 302, Correo 2, Santiago, Chile}

\author{M. Karttunen}
\affiliation{Biophysics and Statistical Mechanics Group, 
Laboratory of Computational Engineering, Helsinki University
of Technology, P.O. Box 9203, FIN-02015 HUT, Finland}

\date{\today}

\begin{abstract}
Collapse and reverse to collapse explosion transition 
in self-gravitating systems are studied by molecular dynamics simulations.  
A microcanonical ensemble of point particles  confined to a spherical box 
is considered; the particles interact via an attractive soft Coulomb
potential. It is observed that the collapse in the
particle system
indeed takes place when the energy of the uniform state is put near or
below the
metastability-instability threshold (collapse energy), 
predicted by the mean-field theory.
Similarly, the explosion in the particle system
occurs when the energy of the core-halo state
is increased above the explosion energy, where according to the mean 
field 
predictions the core-halo state
becomes unstable.
For a system consisting of 125 -- 500 particles, the collapse
takes about $10^5$ single particle 
crossing times to complete, while a typical explosion
is by an order of magnitude faster.
A finite lifetime of metastable states is observed. 
It is also found that the mean-field description 
of the uniform and the core-halo states is exact within
the statistical uncertainty of the molecular dynamics data.

\end{abstract}

\maketitle
\section{Introduction}
\label{sec_intro}

Systems of particles interacting via a potential with attractive
nonintegrable
large $r$ asymptotics, $U(r)\sim r^{-\a}$, $0<\a<3$, and sufficiently 
short-range small $r$ regularization
exhibit gravitational phase transition between a relatively uniform
high energy state and a low-energy state with a core-halo structure
\cite{pr,ki2,ch1,usg,dv,ch,chs,chi}.
Extensive mean-field (MF) studies of the equilibrium properties of 
such systems
\cite{pr,ki2,ch1,usg,dv,ch,chs,chi} revealed that 
in a microcanonical ensemble
during such a transition entropy has to undergo a discontinuous jump
from a state that just ceases to be a local entropy maximum to a state
with the same energy but different temperature,
which is the global entropy maximum. Due to the long-range nature of
gravitational interaction, the MF studies are believed
to provide asymptotically (in the infinite system limit) exact
information about the 
density and the velocity distributions and other thermodynamical parameters 
of the uniform state. The applicability of the MF theory to the
description of the core-halo state is less obvious as the properties
of a core are controlled by the short-range asymptotics of the potential.

Relatively little is  known about how such a 
transition actually occurs, however. 
Youngkins and Miller~\cite{bm2} performed
a Molecular Dynamics (MD) study of a one-dimensional
system consisting of concentric spherical shells. 
Their main emphasis was to check the MF description of the stable and 
metastable states rather than to study the dynamics of the phase transition
itself.
Cerruti-Sola, Cipriani, and Pettini~\cite{pet} studied the
phase diagram of a more realistic 3-dimensional particle
system by using Monte Carlo  and  MD methods.
Their studies again focused on the equilibrium properties
of the system rather than on the dynamics of the transitions. In addition, 
their general conclusions about the second order of the 
gravitational phase transition apparently contradict the MF results
\cite{ki2,usg,dv,chi}. 

Here, we attempt to resolve this contradiction. 
A MF description of the dynamics of collapse in ensembles of
self-gravitating Brownian particles with a bare
$1/r$ interaction based on a Smoluchowski equation was developed by Chavanis
et al 
\cite{chs}. It predicts a self-similar evolution of the central part of density
distribution to a finite-time singularity. However, the
precise nature of the random force and friction terms in
the corresponding Fokker-Plank equation as well as the applicability of the 
overdamped limit used to reduce the Fokker-Plank equation to the Smoluchowski
equation are not entirely understood. A more rigorous approach based
on the Fokker-Plank equation with Landau collision integral was used
by Lancellotti and Kiessling \cite{ki3} to prove 
a scaling property of the central part of the density
profile. The model considered there allows the particles to escape to 
infinity and therefore does not have an 
equilibrium or even a metastable state.

There exists a vast amount of literature on cosmologically- and 
astrophysically- motivated studies of the temporal evolution
of naturally occurring self-gravitating systems 
(see, e.g., Ref.~\cite{cos} and 
references therein). The selection of  systems and their 
initial and final conditions made in such studies are typically
astrophysically-motivated; the considered systems are often too complex 
to make a general
conclusions about the phase diagrams and phase transitions in such systems.

In this paper we present  MD studies of gravitational collapse, and 
reverse to collapse, i.e.,  explosion, transitions in a microcanonical ensemble
of
self-attracting particles. Besides their pure 
statistical mechanical implications, these studies
represent our attempt to bridge a gap between the usually complicated MD and 
hydrodynamic simulations of the realistic
astrophysical systems and the MF analysis of the phase diagram
of simple self-gravitating models.

A system with 
soft Coulomb potential 
$-(r^2+r_0^2)^{-1/2}$ is considered. Such systems are have been studied
using both MF theory (see, e.g., Refs.~\cite{usg,chi}) and simulations
\cite{pet}.
We chose the microcanonical ensemble
as the most fundamental one for the long-range interacting systems.
It has to be noted that the considered system is strongly ensemble-dependent: 
While the nature of the uniform state is the same in both 
microcanonical and canonical ensembles (apart form the difference in their
stability range), the core-halo states and the collapse itself in these 
ensembles
have very little in common with each other \cite{dv,chs}.

\begin{figure}
\includegraphics[width=.45\textwidth]{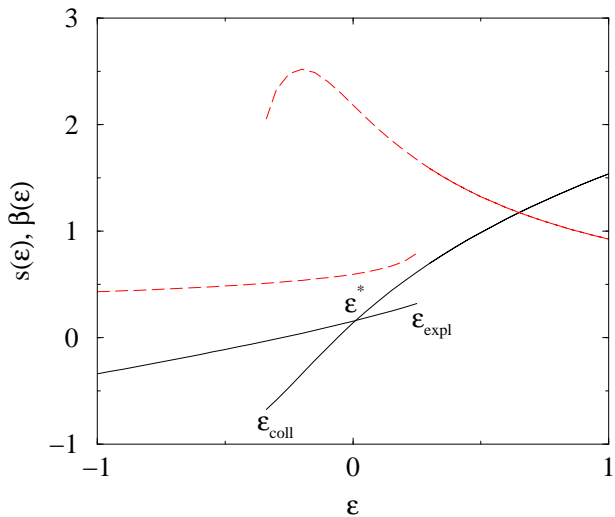}
\caption{\label{fig_mf}
Plots of entropy $s(\e)$ (solid line) and inverse
temperature $\b(\e)=ds/d\e$ (dashed line) vs. energy $\e$
for a system with a gravitational phase transition and a short-range
cutoff.}
\end{figure}

A MF phase diagram of the considered self-attracting microcanonical
system is presented in Fig.~(\ref{fig_mf})
\cite{usg,chi}. 
High- and low-energy branches terminating at the
energies $\e_{coll}$ and $\e_{expl}$ correspond to the uniform and core-halo
states. The energy $\e^*$ where the entropies of the core-halo and uniform
states are equal is the energy of the true phase transition; the
uniform and the core-halo states are metastable in the energy intervals
$(\e_{coll},\e^*)$ and $(\e^*,\e_{expl})$, respectively. However, 
for the phase transition to occur at or near  $\e^*$, a macroscopic-scale
fluctuation with prohibitively low entropy is required.
Consequently, the metastable branches are stable everywhere 
except at the vicinity of $\Delta \e \sim N^{-2/3}$ of their end-points
$\e_{coll}$ and $\e_{expl}$  \cite{ka2,chi}, where $N$ is the number
of particles in the system. Hence it is natural to assume that once the energy 
of the system in the uniform state is set sufficiently near above or below 
$\e_{coll}$, the system
will undergo a collapse to a core-halo state with the same energy and higher
entropy. Similarly, if the energy of the core-halo system is set sufficiently 
near below or above $\e_{expl}$, the system
will undergo an explosion bringing it to a uniform state with the same energy 
and 
higher entropy. Our goal here is to study if and how such collapse and
explosion
proceed in a realistic three-dimensional N-particle dynamical system.

The paper is organized as follows. In the next section we formally introduce
the system, outline the MF analysis  and describe the MD
setup. Then, we present  the simulation results 
for the equilibrium  uniform and core-halo states and compare them to the 
MF predictions. After that we describe and interpret
the observed dynamics of the collapse and the explosion transitions.
A  discussion of the obtained results concludes the paper.

\section{simulation}
We consider a system consisting of $N$ identical particles of unit mass
confined
to a spherical container of radius $R$ with reflecting walls.
The particles interact via the attractive soft Coulomb pair
potential $-(r^2+r_0^2)^{-1/2}$. 
Using a traditional convention for self-gravitating systems in which the 
equilibrium properties of such systems become universal, 
we define rescaled energy $\e$, inverse temperature $\b$, distance $x$,
velocity
$u$, and time $\t$ as
\begin{eqnarray}
\label{def}
\nonumber
\e\equiv E{R\over N^2}\\
\nonumber
x\equiv{r\over R}\\
\b\equiv{N\over RT}\\
\nonumber
u \equiv v \sqrt{R \over N}\\
\nonumber
\t\equiv t {N^{1/2}\over R^{3/2}}.
\end{eqnarray}
The unit of time, often referred to as  crossing time, 
$[t]=\frac {R^{3/2}} {N^{1/2}}$, is obtained by dividing
the unit of length $R$ by the unit of velocity $\sqrt{N/R}$.
This unit of time is also proportional to the period of plasma oscillations
in a medium with the charge concentration $N/R^3$.
As this time unit has purely kinematic origin, we do not expect
the evolution of systems having different $N$ and $R$ to be
universal in time $\t$.
The evolution, assuming that it is collisional, 
is expected to be universal in the relaxation time
$\t_{r}=\t{\ln N\over N}$
\cite{bt}, where the factor $N/ \ln N$
is proportional to the number of crossings a typical particle needs to
change its velocity by a factor of 2 through weak Coulomb scattering events.

The soft core radius $x_0=r_0/R=5\times 10^{-3}$
is chosen to be well below the critical
value $x_{gr}\approx 0.021$, above which the collapse-explosion transition is 
replaced by a normal first-order phase transition \cite{chi}.

The MF theory of the system is described in detail in Ref.\,\cite{usg}.
The  equilibrium velocity distribution is Maxwellian and isothermal, while 
the equilibrium (saddle point) density profile $\r({ x})$ 
corresponding a stable or a metastable state, is a spherically 
symmetric solution of the integral
equation (\ref{extr}).  This equation replaces
the Poisson-Boltzmann differential equation for the self-consistent
potential (see, for example, Ref.\,\cite{pr}) 
since the interparticle interaction considered here is not pure Coulombic.
\begin{figure}
\includegraphics[width=.45\textwidth]{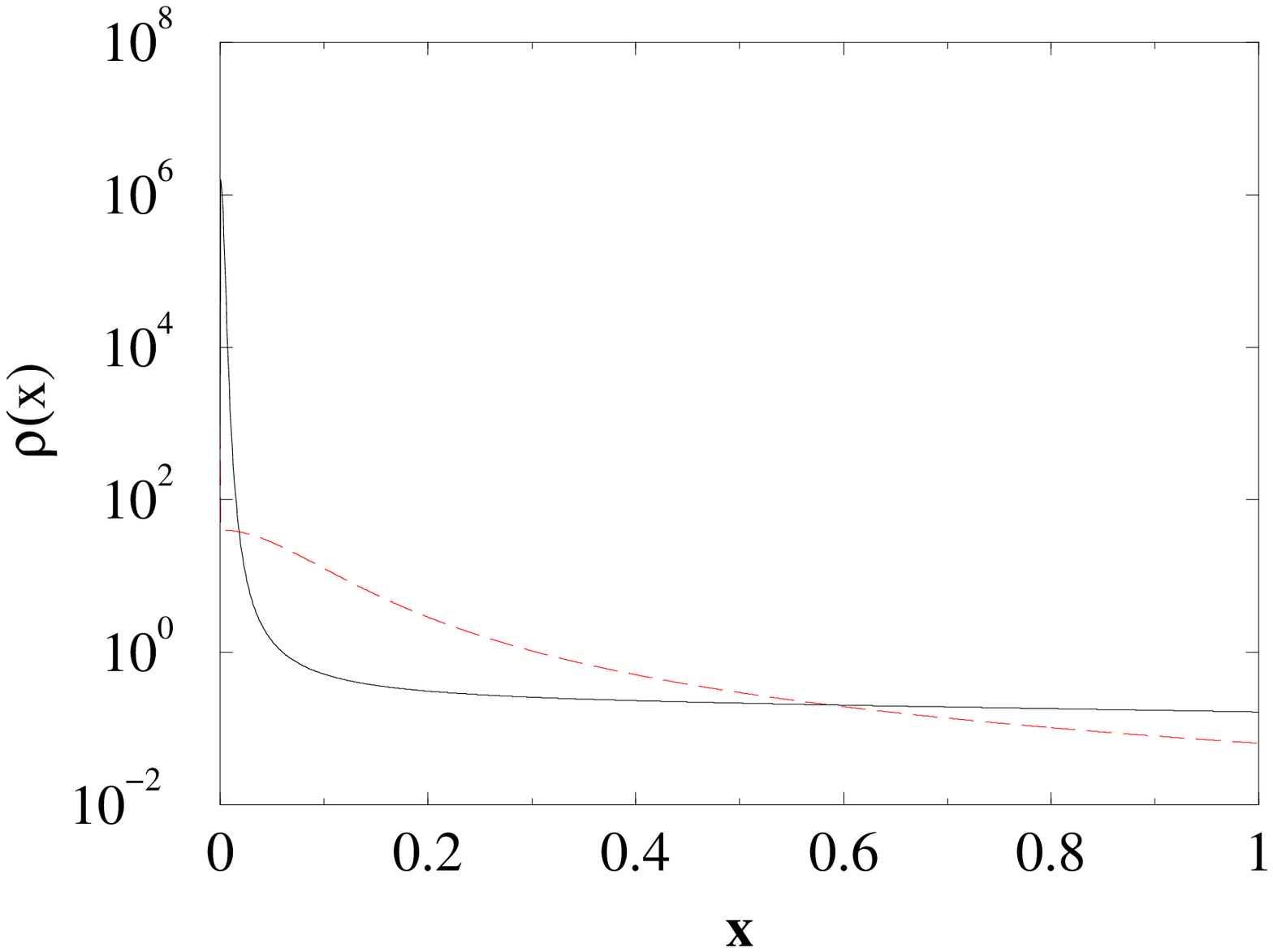}
\caption{\label{fig_rho}
MF density profiles $\r_u({x})$  
of a uniform state (dashed line) and $\r_{c-h}({x})$
of a core-halo state (solid line)
for $\e=\e_{coll}$ }
\end{figure}
\begin{eqnarray}
\label{extr}
\nonumber
\r({\bf x})=\r_0 F[\r(.), {\bf x}]\\
\nonumber
F[\r(.),{\bf x}]=\exp\left [\b \int {\r({\bf x}')\over
\sqrt{({\bf x}-{\bf x}')^2+x_0^2}}d^3{\bf x}'\right ]\\
\b={3\over 2} \left[\e + {1\over 2}\int \int{\r({\bf x}_1) \r({\bf x}_2)
\over \sqrt{({\bf x}_1-{\bf x}_2)^2+x_0^2} }d^3{\bf x}_1 d^3{\bf x}_2
\right ]^{-1}\\
\nonumber
\r_0=\left\{\int F[\r_s(.), {\bf x}]d^3{\bf x}\right\}^{-1}
\end{eqnarray}
The equilibrium density profile $\r({x})$ obtained from this 
equation is 
then used to calculate
the entropy and the pressure.

The MF phase diagram of the system is presented in Fig.~\ref{fig_mf}. 
The collapse and
explosion energies
are $\e_{coll}\approx -0.339$ and $\e_{expl}\approx 0.267$. Examples of the
uniform and the core-halo density profiles for $\e=\e_{coll}$ are shown in
Fig.~\ref{fig_rho}.

In the MD simulations we consider a system consisting of
$N=125$ -- 500 particles in a  spherical container of the radius $R=1$.
All interparticle forces are calculated directly 
at each time step $dt$. This is done to avoid any mean-field-like
effects inevitably present in any truncated
multipole or Fourier potential  expansion.
The particle velocities and coordinates are updated according to
the velocity-Verlet algorithm which is symplectic and reversible. 
 
The system is initialized by randomly distributing particles according to a
spherically symmetric density profile; typically
the appropriate MF density profile $\r(x)$ was used. The potential energy
($U$) of the initial configuration is calculated, and the target kinetic energy
$E_k=E-U$ is determined. The particle
velocities are randomly generated from some (usually Maxwell) distribution
with the appropriate square average. Finally, the deviation of the total
energy from its target value, caused by a stochasticity in velocity
assignment, is determined, and the velocities are rescaled to
fine-tune the total energy.  
Due to the isotropicity of the random velocity assignment,
we have always obtained the states with sufficiently low total angular 
momentum
which collapsed to single-core states rather than to binaries \cite{grv}.

To implement the reflective boundary condition, at each time step the
normal components $v_{\perp}$ of the velocities of all
particles which had escaped from the container were reversed.
Values of the normal components were stored to
evaluate the pressure on the wall $P$. 
\begin{equation}
\label{P}
P(t)=\frac{\sum_{t'=t-t''/2}^{t'=t+t''/2}v_{\perp}(t')}{2\pi R^2 t''}
\end{equation}

During each simulation run we measured such characteristics
as the kinetic energy $\e_{k in}=\frac{3}{2\b}$, the virial 
variable
$\s$ (dimension of energy) quantifying deviations
from the virial theorem,         
$\s \equiv \e +  \e_{kin} -\frac{3 PV R}{N^2}$ 
(where $P$ is the pressure on the wall,
$V=4\pi R^3/3$  is 
volume of the container, and the factor $N^2/R$
rescales the volume-pressure term to the unit of energy introduced in
(\ref{def})), ratio of the velocity moments, 
$ \frac{19 \langle v^2 \rangle ^2}{
5 \langle v^4 \rangle }$ (which should be 1 for the Maxwell distribution), and 
the number of particles in the core $N_c$ of the prescribed radius $x_c$. 
For the last measurement we count the number of particles $N_i$ that are 
within $x_c$ from the $i$th particle and find the particle which has
the largest $N_i$.

In addition, we measured the histograms of the velocity distribution
and radial distribution functions, $W(u)$ and $C(x)$, respectively. The
latter was defined as the number of particles in the spherical layer
of the radius $x$ around each particle, normalized by the volume of such layer,
disregarding the nonuniformity of the system and the boundary effect.

The measurements of the ''scalar'' quantities such as energy, kinetic energy,
pressure, and velocity distribution moments were taken in time intervals
$\t_{meas}$, which were selected sufficiently long to avoid
measuring the unchanged configuration repeatedly and sufficiently
short not to miss the important details of the system evolution.
We usually pick $\t_{meas}$  of the order of the uniform density sphere 
crossing time $\t_{cross}^u=\pi$, which is a 
half period
of the oscillation of a particle released with zero velocity at the container 
wall. 
The histogram data, such as the velocity  distribution and the 
radial distribution 
functions,  was incremented at each $\t_{meas}$ and accumulated over
a longer time period $\t_{hist}$, $\t_{hist}\sim 10$ -- $10^3 \times
\t_{meas}$.

Our attempts to resolve the high-density part of the 
radial density profile of the 
system turned out to be fruitless due to the strong fluctuations in the 
positions of this part. This fluctuations result in smearing the
central peak in both core-halo and low energy uniform states. 
Considering the center of mass  system of 
reference does not resolve this difficulty, as, despite being dense, 
the core typically contains only 10 -- 20\% of the total  system mass 
(see below)
and the positions of the core and the center of mass of 
the system do not usually coincide.

To control the quality of the simulation, we monitored the total energy $\e$
and the total angular momentum $L$. 
We selected timestep $dt$ small enough to keep
the total energy variation within 0.01\% of its initial value, 
usually we used $dt =10^{-5}$, or in rescaled units, $d\t \sim 10^{-4}$. 
For such time steps,
the relative deviation of the angular momentum was within $10^{-14}$.


All the measurements below are presented in the rescaled dimensionless units
as defined in Eq.~(\ref{def}).

\section{uniform and core-halo equilibrium states: comparison to the MF}

To check our simulation procedure and possibly resolve the apparent 
contradiction
between the MF and the particle simulation results \cite{pet},
we first considered the system in what we expected to be a stable or a 
metastable states 
far away from a transition point. Since we were interested in the 
equilibrium
properties, we were initiating the MD systems
according the corresponding MF predictions.
It meant that the density profiles were seeded according to the
MF profiles and the velocities were assigned according to the Maxwell
distribution. 
We observed that the MF density initiation virtually eliminates the
transitory period, while the method of velocity assignment was practically 
unimportant, provided that it gave the correct value for the total kinetic
energy. For example, it takes  a system initialized with a flat 
$W({\bf u})=const$ about $\t\sim \t_r$ to evolve to the Maxwell
distribution.
   
\begin{figure}
\includegraphics[width=.45\textwidth]{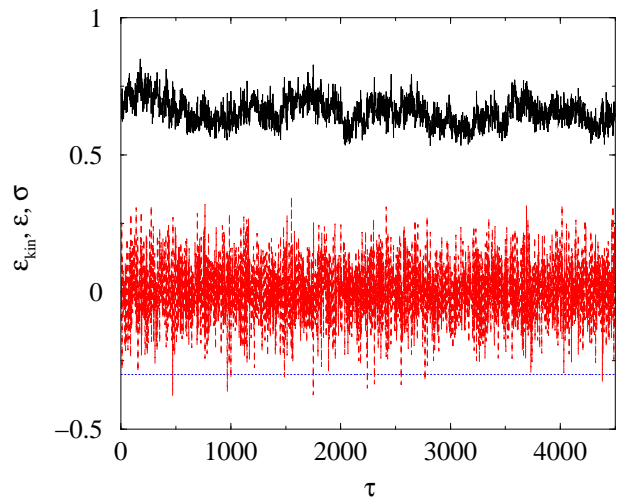}
\caption{\label{fig_td}
Plots of time dependence of kinetic energy $e_{kin}$ (solid line), virial
variable
$\s$ (dashed line), and total energy $\e$
(dotted line) for a uniform system of $N=250$ particles
at $e=-0.3$. }
\end{figure}

A typical plot of the steady state time dependence of the kinetic energy,
virial variable, and the total energy is presented in Fig.~\ref{fig_td}.

The comparison between the MD measurements and the MF results for the 
uniform and 
core-halo states is presented in Tables 
\ref{tab_un} and \ref{tab_ch} and reveals a perfect agreement between
these two sets of data.
\begin{table}
\caption{Equilibrium MD and MF results for a uniform 
state for $\e=-0.3$, $N=250$, and $0\leq \t \leq 5000 $}
\begin{ruledtabular}
\begin{tabular}{ccc}
\label{tab_un}
 & MD & MF \\
\hline
$\e$ & $-0.3 \pm 5\times 10^{-7}$   & -0.3 \\
$\e_{kin}$ &$0.66 \pm 0.05$   & 0.644 \\
$\s$ &$0\pm 0.03$  &0.012 \\
$ 19 \langle v^2 \rangle ^2 /
5 \langle v^4 \rangle $ &$1.01\pm0.04$ & 1 \\ 
\end{tabular}
\end{ruledtabular}
\end{table}
\begin{table}
\caption{Equilibrium MD and MF results for a core-halo 
state at $\e=-0.339 $, $N=250$, and $0\leq \t \leq 1500 $}
\begin{ruledtabular}
\begin{tabular}{ccc}
\label{tab_ch}
 & MD & MF \\
\hline
$\e$ &$-0.3392 \pm 2\times 10^{-4}$   & -0.339 \\
$\e_{kin}$ &$2.9  \pm 0.1$ & 2.94 \\
$\s$ & $-1.5 \pm 0.1$ &-1.46 \\
$ 19 \langle v^2 \rangle ^2 /
5 \langle v^4 \rangle $ & $0.99 \pm 0.03$ & 1\\
$N_{core}$&$48\pm 2$&47.6\\
\end{tabular}
\end{ruledtabular}
\end{table}
To obtain the expression for the MF virial variable 
\begin{equation}
\label{vir}
\s_{MF}=\e+\e_{kin}(1-8\pi\r(1)/3)
\end{equation}
we write for the  
pressure at the container wall $P=2\rho(x=1)\e_{kin}/3$ 
implying that the system is isothermal. Since the interparticle 
potential is not pure 
Coulombic, the virial variable is non-zero. 
The difference is especially prominent
for the core-halo states where more particles ''probe'' the short-range
part of the potential.

\begin{figure}
\includegraphics[width=.45\textwidth]{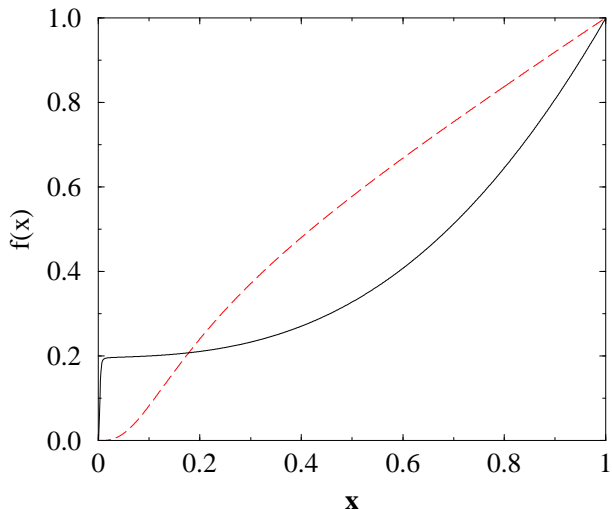}
\caption{\label{fig_irho}
Integrated MF density profiles $f_u({x})$  
of a uniform state (dashed line) and $f_{c-h}({x})$
of a core-halo state (solid line)
for $\e=\e_{coll}$ }
\end{figure}

To evaluate the core radius and number of core particles
of the core-halo system, 
we considered an integrated MF density profile,
$f(x)=\int_0^x 4 \pi y^2 \r(y) dy$
(see Fig.~\ref{fig_irho}).
As it follows from the Figure, the MF core-halo state
indeed contains a distinct core with a sharp boundary of the radius
$x_c\approx10^{-2}$ relatively insensitive to the energy
in the range we considered, $|\e|<0.5$. Using this MF
core radius, we located cores in the MD core-halo
systems which contained very similar to the MF cores number of particles
(see Table~\ref{tab_ch}).
Using smaller core radius resulted in significant reduction in the number
of observed core particles.
A reasonably small 
over-estimation of the core radius did not affect the results of the 
MD measurements: we observed that 
even in the sphere twice the core radius the number of particles
is only marginally  (at most by 8\%) larger than in the core.

To check if the system has more than one core, we performed search for
the second-largest core of the same radius $x_c$. We looked for
a largest group of particles which are within $x_c$ from a single particle 
with none of these particles belonging to the first, largest core.
We never observed the second-largest core containing more than 2 particles; 
most of the time it contained only a single one.

\begin{figure}
\includegraphics[width=.45\textwidth]{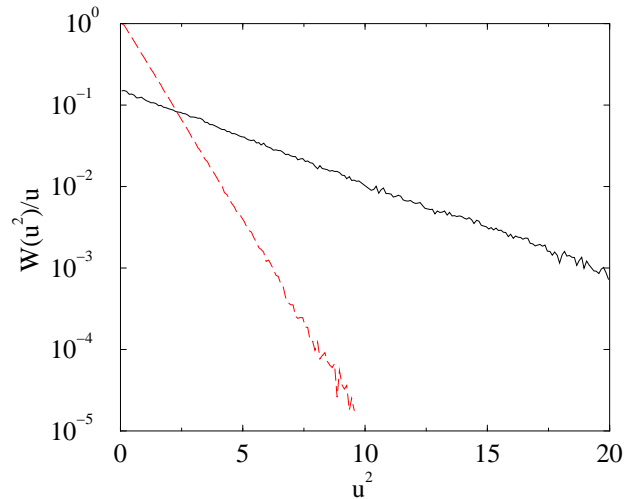}
\caption{\label{fig_vel}
The MD 
velocity distribution functions $W(u)$ of a
core-halo state with $\e=-0.339$
(solid line) and a uniform state
with $\e=-0.3$ (dashed line). In both cases $N=250$.}
\end{figure}

In Fig.~\ref{fig_vel} we present the MD velocity distribution functions 
$W(u)$ for a core-halo and uniform states; shown $W(u)$ confirm the MF 
prediction for the Maxwellian form of these distributions.

\begin{figure}
\includegraphics[width=.45\textwidth]{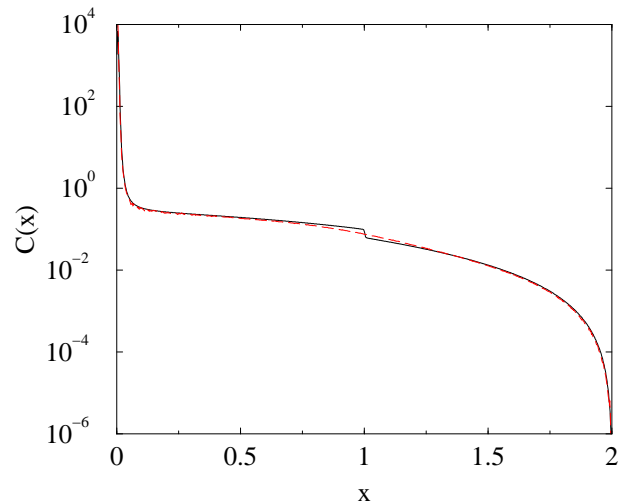}
\caption{\label{fig_rdf}
MF (dashed line) and MD (solid line)
radial distribution functions $C(x)$ of a
core-halo state with $\e=0.25$. The step
at $x=1$ in the MF $C(x)$ is caused by the localization
of the core exactly at $x=0$ and a sharp boundary of the container.}
\end{figure}

As we mentioned in the previous section, we were unable to resolve
the high-density part of the radial density profile due to the 
core motion. However, an indirect comparison between the radial distribution
of particles in the MF and MD was made using the
radial distribution function. The MF radial distribution function
$C_{MF}(x)$ was computed as
\begin{equation}
\label{rdf}
C_{MF}(x)= \frac{1}{4\pi x^2}\int\r({\bf x}')\r({\bf x + x}')d
{\bf x}'.
\end{equation}
The good agreement between the MF and the MD
radial distribution functions is illustrated in Fig.~\ref{fig_rdf}.
This indicates that the mutual distribution of particles is correctly 
predicted by the MF theory.

To summarize, for all the quantities considered, 
we observed no systematic deviations between the MF theory and the MD data.

\section{Collapse}
According to the MF theory, if the energy of the uniform state
becomes lower than $\e_{coll}\approx-0.339$, the system should undergo
a collapse to a core-halo state. To study the collapse, we considered several
uniform systems with the energies ranging between $\e=-0.5$ and $\e=-0.3$. 
The systems were
initialized according to the MF density distributions. For systems
with $\e<\e_{coll}$ the particles were distributed
according to the MF density profile for $\e_{coll}$. 

\begin{figure} 
\includegraphics[width=.45\textwidth]{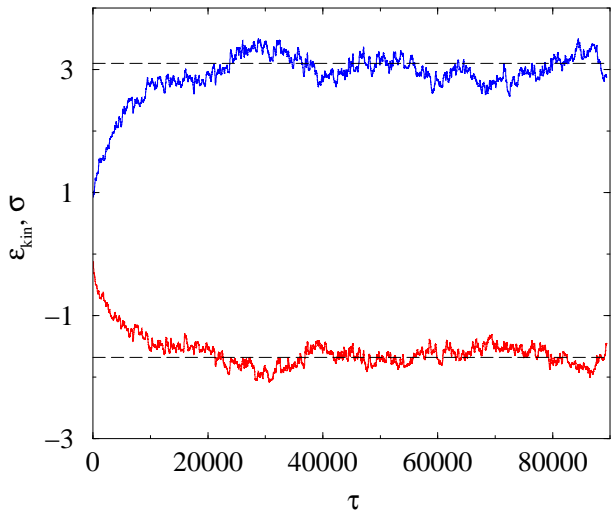}
\caption{\label{fig_coll}
Time dependence of the kinetic energy $\e_{kin}$ (top) and the virial variable 
$\s$ (bottom) of the
collapsing uniform state with $\e=-0.5$ and $N=125$.
The dashed horizontal lines indicate the equilibrium values of $\e_{kin}$ and 
$\s$ of the corresponding core-halo state. The data is averaged
over $\d \t=100$ time intervals.}
\end{figure}

\begin{figure}
\includegraphics[width=.45\textwidth]{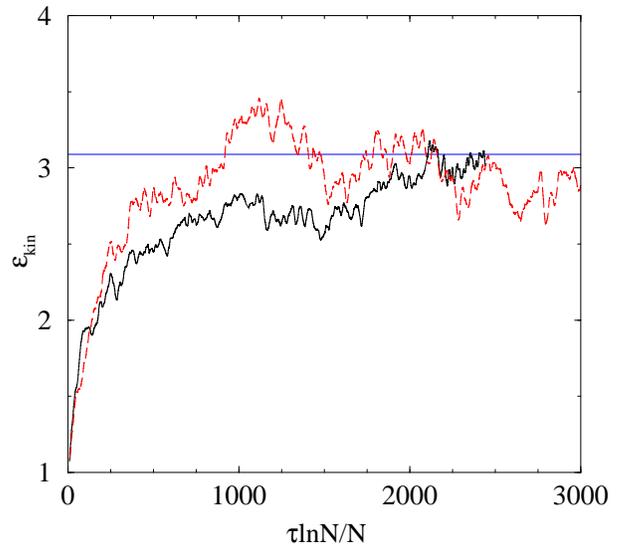}
\caption{\label{fig_N}
Collapse in systems with $\e=-0.5$ and 
different numbers of particles,
$N=125$ (dashed line) and $N=250$ (solid line), shown
in relaxation time units, $\t\ln N/N$ \cite{bt}. 
Horizontal line shows
the kinetic energy of the target core-halo state. The data is averaged
over $\d \t=100$ time intervals.}
\end{figure}

In a perfect agreement with the MF theory, a uniform state with
$\e<\e_{coll}$
undergoes a gradual transition to a core-halo state with a 
typical timescale of $\t_{coll}\sim 10^4$ for $N=125$ -- 250 particles. 
An example of the time dependence of the kinetic energy and the virial variable
for 
a collapsing system is shown in Fig.~\ref{fig_coll}. 
We observe that if the number of particles is increased
but the rescaled energy $\e$ is kept fixed, it takes generally 
longer time for the collapse to be
complete.  Our results (see Fig.~\ref{fig_N}) qualitatively confirm that
the characteristic time for the full collapse scales as $\t_r$ \cite{bt}.
A quantitative study of the dependence of the collapse dynamics on 
the number of particles requires much faster simulation code, however.

\begin{figure}
\includegraphics[width=.45\textwidth]{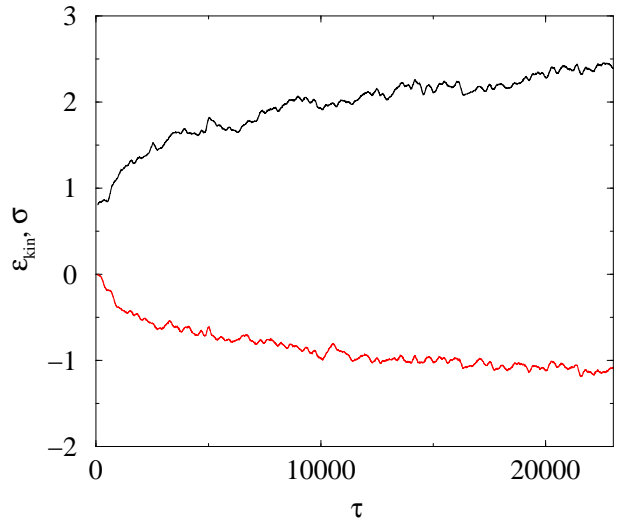}
\caption{\label{fig_ec1}
Plots of the kinetic energy $\e_{kin}$ (top) and the virial variable $\s$
(bottom) 
vs time $\t$ for a system 
with $\e=\e_{coll}\approx-0.339$ 
and $N=250$.The data is averaged
over $\d \t=100$ time intervals.}
\end{figure}

In the above examples, the energy was set  to $\e=-0.5$ which is 
well below $\e_{coll}\approx-0.339$, and as a consequence 
the collapse started immediately at $\t=0$ in all simulation runs.
If the system energy is  $\e_{coll}$, the noticeable
increase in kinetic energy and decrease of the virial variable, 
characteristic for 
collapse, start not exactly at $\t=0$ but with a small delay 
(Fig.~\ref{fig_ec1}) which varies from run to run from almost zero
to about $\t\approx 1500$. 
This indicates that the MD system is able to overcome the 
metastability at or near $\e_{coll}$.  The 
observed uncertainty is likely due to the  relatively small number of particles.

\begin{figure}
\includegraphics[width=.45\textwidth]{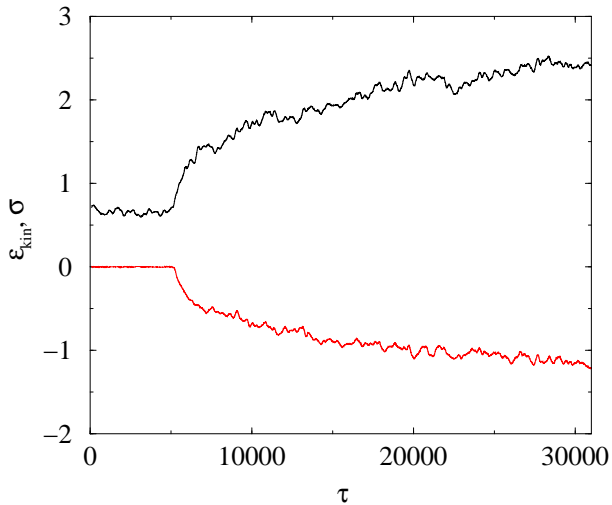}
\caption{\label{fig_met}
Plots of the kinetic energy $\e_{kin}$ (top) and virial variable $\s$ (bottom) 
vs time $\t$ for a system 
with $\e=-0.3$ 
and $N=250$. The data is averaged
over $\d \t=100$ time intervals.}
\end{figure}

As we increase the energy above $\e_{coll}$, the stability of the uniform  
state
increases which results in a longer lifetime of such state with respect to
collapse. In Fig.~\ref{fig_met}, an evolution of a system with $\e=-0.3$
is shown. The system stays in the 
uniform state for about $\d \t \approx 5000$ 
before the collapse starts, after which the evolution proceeds qualitatively
similar to the collapses in systems with lower energies. 

\begin{figure}
\includegraphics[width=.45\textwidth]{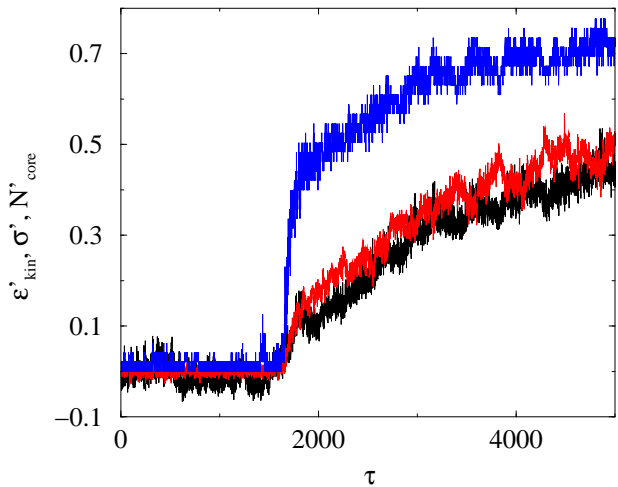}
\caption{\label{fig_ec}
Plots of the relative values of
(from top to the bottom) number of core particles $N'_{core}(\t)$,
virial variable $\s'_{kin}(\t)$, and kinetic energy  $\e'_{kin}(\t)$
for the system with $\e=-0.339$ and $N=250$.}
\end{figure}

To compare the temporal evolution of the kinetic energy, virial variable,
and  the number of core particles,
the relative variables  $\e'_{kin}(\t)$, $\s'_{kin}(\t)$, and
$N'_{core}(\t)$, all defined as  
$\e'_{kin}(t)=[\e_{kin}(t)-\e_{kin}(u)]/[\e_{kin}(c-h)-\e_{kin}(u)]$,
are plotted  in  Fig.\ref{fig_ec}.
The values  $\e_{kin}(u)$ and $\e_{kin}(c-h)$ correspond to
the uniform and core-halo states in equilibrium. 

Figure~\ref{fig_ec} indicates that during the initial stages of collapse the 
core grows faster than the kinetic energy
and the virial variable.
In addition, one can notice large
reversible fluctuations in the number of core particles (the core
grows up to 12\% of its final value and then disappears) that are not matched
by comparable scale fluctuations in the kinetic energy or virial variable.
All these observations suggest
that the density evolution 
causing the core formation plays the
leading role in the process of collapse while the relaxation
of kinetic energy follows.
Once the collapse has started, the core grows to about a half of its final 
size in
only $\d\t_{core}\sim 10^3$ for systems with $N=125$ -- 500 particles, while
the changes in kinetic energy during this interval of time are small.
After this rapid initial stage the system relaxes more slowly,
and finally after $\t_{coll}\sim 10^5$ reaches the equilibrium 
core-halo state.
Our observations strongly suggest that the growth of the core takes place
through
a sequential absorption of single particles rather than through hierarchical 
merging of smaller cores: We never detected
other cores containing  more than two particles.
Although the kinetic energy relaxation trails behind the the core
formation, the velocity distribution function
remains Maxwellian throughout the whole evolution 
with the temperature corresponding to the corresponding
value of the kinetic energy. This is caused by 
the fast velocity relaxation ($\t_{vel} \leq 1$)
as discussed in the previous section.

\section{Explosion}
It is natural to assume that if a system exhibits a collapse, it should
also exhibit an explosion which is the reverse to the collapse transition.
According to the MF theory, such explosion should take place
when the core-halo state becomes unstable, i.e., when $\e \geq \e_{expl}\approx
0.267$.
To check this prediction, we initialized the MD system according to the MF
equilibrium 
core-halo state and followed its evolution. As in the study of the collapse,
for initial states with $\e > \e_{expl}$ we used the MF density profiles
of the highest energy locally stable state, i.e. of the state with 
$\e = \e_{expl}$.

\begin{figure}
\includegraphics[width=.45\textwidth]{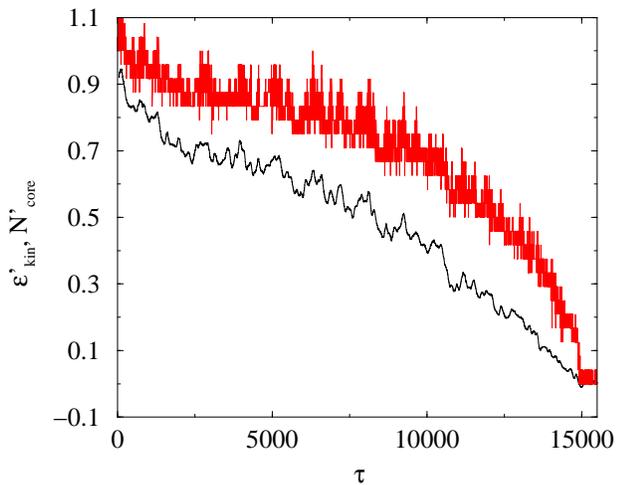}
\caption{\label{fig_expl}
Plots of the kinetic energy $\e'_{kin}(\t)$ (top) and relative number of core 
particles $N'_{core}(\t)$
(bottom) (defined as in Fig.~\ref{fig_ec})
vs time $\t$ for a system 
with $\e=0.5$ 
and $N=250$. The kinetic energy is averaged
over $\d \t=100$ time intervals. 
}
\end{figure}

\begin{figure}
\includegraphics[width=.45\textwidth]{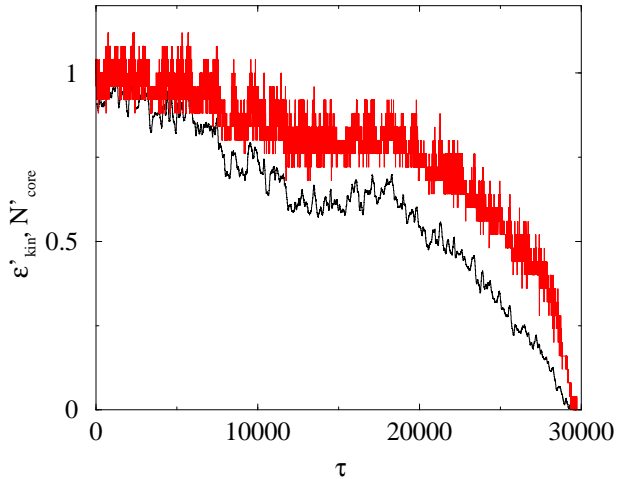}
\caption{\label{fig_expla}
Same as in Fig.~\ref{fig_expla} but for $\e=0.4$. 
}
\end{figure}

We observe that a system with sufficiently high energy, such as 
$\e =0.5$ in Fig.~\ref{fig_expl} or $\e =0.4$ in Fig.~\ref{fig_expla},
indeed undergoes an explosion which brings
it to the uniform equilibrium state. During such an explosion, the
state variables such as kinetic energy and virial variable continuously change 
from
their equilibrium core-halo state values to the uniform state ones, and
the core gradually sheds  particles until only one particle is left.
The main features of an explosion (Figs.~\ref{fig_expl}~and~\ref{fig_expla}) 
resemble those of a time-reversed collapse. 
The kinetic energy evolves relatively uniformly, while
the number of core particles changes only slightly during the first stages of 
evolution and rapidly decreases at the final stages.
In the example presented in  Fig.~\ref{fig_expl}, the explosion is complete
after the time $t_{expl}\approx 15000$, which is noticeably less than the time
for a  collapse $t_{coll}\approx 10^5$ (see Fig.~\ref{fig_N}) 
for a system having the same number of 
particles ($N=250$). 
However, the latter is rather vaguely defined due to larger
fluctuations in a core-halo than in a uniform state. 

\begin{figure}
\includegraphics[width=.45\textwidth]{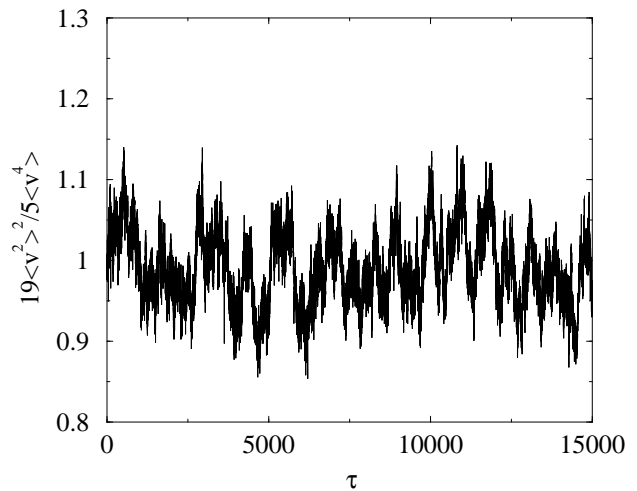}
\caption{\label{fig_velm}
Plot of  he ratio of the moments
of velocity distribution,  $19 \langle v^2 \rangle ^2 /
5 \langle v^4 \rangle$, vs.time $\t$ for a system with
$\e=0.5$ and $N=250$.}
\end{figure}

Similarly to a collapse, the system remains thermalized 
in the velocity space during an explosion. The velocity distribution remains
Maxwellian 
throughout the 
evolution with the temperature corresponding to the current value of kinetic
energy. As an illustration, Fig.~\ref{fig_velm} 
shows the ratio of the moments
of velocity distribution,  $19 \langle v^2 \rangle ^2 /
5 \langle v^4 \rangle$, which should be 1 for a Gaussian distribution. 

However, as it is evident from a comparison between the Figs.~\ref{fig_expl}
and \ref{fig_expla}, as $\e$ gets closer to  $\e_{expl}$ the explosion takes
longer to initiate. We have observed, that even for $\e=0.3$, which is 
noticeably 
larger than $\e_{expl} \approx 0.267$, the explosion does not happen during
the first $\t=30000$ of evolution. This suggests that either the MF 
value for  $\e_{expl}$ is incorrect, or during the incitation of the system
we somehow prepare the system not exactly in the equilibrium (metastable) 
core-halo state. If the latter is the case, a deviation from the equilibrium
most probably takes place in the core, as because of its compactness, its
equilibration with the rest of the system may take a rather long time.
Using the current MD setup, we were unable to determine a reason for this
apparent discrepancy. 

\section{CONCLUSION}

In the previous sections we have presented the following molecular dynamics 
results for the self-attracting systems with soft Coulomb potential: 
\newcounter{ref}
\begin{list}{\arabic{ref}.}{\usecounter{ref} \itemsep=0pt
\leftmargin=1cm}
\item A collapse from a uniform to a core-halo state was observed.
The timescale for the collapse in systems consisting of 125 -- 500 particles
is of order of $10^5$ crossing times and 
is by the same factor longer than the timescale of the 
velocity relaxation. The collapse starts with a fast growth of a core
via absorption of single particles and continues with more gradual relaxation
towards an equilibrium core-halo state. Metastable states exhibit a finite 
lifetime before collapsing. 
\item A reverse to collapse, i.e., an explosion transition from a core-halo to a

uniform state was observed. The explosion time is considerably shorter than
the collapse time, being of the order of  $10^4$ crossing times
(125 -- 500 particles). An explosion resembles a time-reversed
collapse; the core decrease, which happens by shedding individual particles, 
is trailing the kinetic energy evolution till the last stages, when the core
rapidly disappears.
\item Such molecular dynamics characteristics of the equilibrium or 
the metastable 
uniform and core-halo states as kinetic energy, wall pressure,
number of core particles, particle-particle radial distribution function
and velocity distribution function,
are found to be equal within the statistical uncertainty of the molecular
dynamics measurements to the corresponding mean field predictions.
\end{list}

The long collapse time observed in our simulations appears to be an
explanation for the apparent discrepancy between the phase diagram presented
in \cite{pet} and the mean field phase diagram (see, for example, \cite{chi}).
The relaxation time allowed in \cite{pet} before the measurements of what
was considered to be a steady state, $t_{rel}=3N/|EN|^{3/2}$, which is 
apparently equivalent to $\t_{rel}<1$, is by far insufficient for
a system to collapse. Therefore, the discontinuities in caloric
curves $\beta$ vs $\e$, typical for collapse and explosion gravitational
transitions, were not observed in \cite{pet}.

Although we considered systems only with the soft Coulomb potential,
we speculate that a likewise similarity between the mean field and molecular
dynamics equilibrium properties of the core-halo state 
exists for all ''soft'' long-range
(like a Fourier-truncated Coulomb) potentials. This is so because
all soft potentials are effectively longer-ranged than the bare
Coulomb one. However, the core-halo state in the system with a ''harder''
short-range cutoff may have completely different properties from the one 
considered above, and its mean field theory may be inadequate.
As for the uniform states, their properties are virtually independent on the
nature of the cutoff (see, for example, \cite{chi}) and their mean-field
description is universally correct.

The main goal in the paper was to check the existence of collapses and 
explosions and the validity of the mean field data for the
 self-gravitating systems
with short-range cutoff. For this goal one or few molecular dynamics runs 
for each considered system were sufficient. However, to be able to
study the dynamical features of collapses and explosion in more detail and
to compare the simulations results to various theoretical models, one
needs to study the relaxation 
averaged over many initial configuration. For example, an interesting 
question 
is whether a collapse (or an explosion) indeed consists of two stages;
the first fast stage of collisionless ''violent relaxation''
with particle number-independent rate, and the slower second stage
of soft collisional relaxation with characteristic time $\t_r$ 
(see, for example, \cite{bt} and references therein).
Another important question is to resolve the apparent contradiction
between the mean field prediction for $\e_{expl}$ and the molecular dynamics
observations, outlined at the end of the previous section.
Such studies require a more efficient
computation code. 
The main improvement possibly coming from a better force
calculator that may include various mean-field-like potential expansions,
which are qualitatively justified by this study.
We leave this for the future research.

\section{acknowledgments}

The authors are thankful to  P.-H.~Chavanis and  E.~G.~D.~Cohen for
helpful and inspiring discussions  and gratefully acknowledge the support
of Chilean FONDECYT under grants 1020052 and 7020052.
M.\,K. would like to thank the Department of Physics at 
Universidad de Santiago for warm hospitality.

\end{document}